\documentclass[preprint]{aastex} 

\shorttitle{Gliese 436b's Eccentricity} 
\shortauthors{Batygin et al.} 

\newcommand{\be}{\begin{equation}}
\newcommand{\ee}{\end{equation}}

\def\lta{\,\raise 0.3 ex\hbox{$ < $}\kern -0.75 em
 \lower 0.7 ex\hbox{$\sim$}\,}
\def\gta{\,\raise 0.3 ex\hbox{$ > $}\kern -0.75 em
 \lower 0.7 ex\hbox{$\sim$}\,}

\begin{document}
 
\title{A Quasi-Stationary Solution to Gliese 436b's Eccentricity}  
\author{Konstantin Batygin$^{1,2}$, Gregory Laughlin$^{1}$, Stefano Meschiari$^{1}$, Eugenio Rivera$^{1}$, Steve Vogt$^{1}$, Paul Butler$^{3}$ } 

\affil{$^1$UCO/Lick Observatory, University of California, Santa Cruz, CA 95064} 
\affil{$^2$Division of Geological and Planetary Sciences, California Institute of Technology, Pasadena, CA 91125} 
\affil{$^3$Department of Terrestrial Magnetism, Carnegie Institution of Washington, Washington DC 20015} 

\email{kbatygin@gps.caltech.edu}
 
\begin{abstract} 

We investigate the possibility that the large orbital eccentricity of the transiting Neptune-mass planet Gliese 436b is maintained in the face of tidal dissipation by a second planet in the system. We find that the currently observed configuration can be understood if Gliese 436b and a putative companion have evolved to a quasi-stationary fixed point in which the planets' orbital apses are co-linear and in which secular variations in the orbital eccentricities of the two planets have been almost entirely damped out. In our picture, the two planets are currently experiencing a long-period of gradual orbital circularization. Specifically, if Gliese 436b has a tidal $Q\sim300,000$, similar to both the Jovian $Q$ and to the upper limit for the Neptunian $Q$, then this circularization timescale can be of order $\tau\sim 8 \,{\rm Gyr}$ given the presence of a favorably situated perturber.  We adopt an octopole-order secular theory based on a Legendre expansion in the semi-major axis ratio ${a_1}/{a_2}$ to delineate well-defined regions of ($P_{\rm c}$, $M_{\rm c}$, $e_{\rm c}$) space that can be occupied by a perturbing companion. This description includes the leading-order effects of General Relativity, and retains accuracy for perturbing companion planets that have high eccentricity. We incorporate the evolutionary effect of tidal dissipation into our secular model of the system, and solve the resulting initial value problems for a large sample of the allowed configurations. We find a locus of apsidally aligned configurations that are (1) consistent with the currently published radial velocity data, (2) consistent with the current lack of observed transit timing variations,  (3) subject to rough constraint on dynamical stability, and which (4) have damping time scales consistent with the current multi-Gyr age of the star. We then polish the stationary configurations derived from secular theory with full numerical integrations, and compute the transit timing variations and radial velocity half-amplitudes induced by the resulting configurations. We present our results in the form of candidate companion planets to Gliese 436b. For these candidates, radial velocity half-amplitudes, $K_{c}$, are of order $3\,{\rm m s^{-1}}$, and the maximum amplitude of orbit-to-orbit transit timing variations are of order $\Delta t=1\, {\rm s}$ to $\Delta t=5\,{\rm s}$ . For the particular example case of a perturber with orbital period, $P_{c}=40\,{\rm d}$, mass, $M_{c}=8.5\, M_{\oplus}$, and eccentricity, $e_{c}=0.58$, we confirm our semi-analytic calculations with a full numerical 3-body integration of the orbital decay that includes tidal damping and spin evolution. Additionally, we discuss the possibility of many-perturber stationary configurations, utilizing modified Laplace-Lagrange secular theory. We then perform a proof-of-concept tidally dissipated numerical integration with 3 planets, which shows the system approaching a triply circular state. 

\end{abstract}


\section{Introduction} 

In the years since the discovery of the planetary companion to 51 Peg
(Mayor \& Queloz 1995), considerably {\it fewer} than 1,000 additional extrasolar planets
have been detected.\footnote{For an up-to-date listing of the current galactic planetary
census, see http://www.exoplanet.eu} Because the aggregate of known planets is still
quite limited, the discovery and characterization of additional planets orbiting specific
individual stars, remains a topic of considerable interest. 

The detection of Gliese 436b by Butler et al. (2004) marked the first discovery of a 
Neptune mass planet since LeVerrier's discovery of Neptune itself in 1846. In May, 2007,
Gliese 436b was discovered to be observable in transit (Gillon et al. 2007)
which has sparked an intense interest in this remarkable planet. The transits
resolve the $\sin(i)$ degeneracy, and fix the mass of the planet at $m_{b}=23.17 \pm 0.79 \,m_{\oplus}$ (Torres 2007). The period of the planet is $P=2.643904 \pm .000005$ days (Bean et al. 2008)
and it orbits an $M_{\star}=0.452^{+0.014}_{-0.012}\, M_{\odot}$ red dwarf primary (Torres 2007). To date, no timing variations have been observed for the transits of Gliese 436b, with strict orbital periodicity maintained to a level of at least $\sim 7$ seconds (Pont et al. 2008).

The radius, $R_{b}=4.22^{+0.09}_{-0.10} \, R_{\oplus}$ (Torres 2007) of Gliese 436b
indicates that it necessarily contains several earth masses of hydrogen and helium
in its outer layers. It probably resembles Neptune in overall composition
(see, e.g. the models of Fortney et al. 2007). Measurement of the secondary eclipse depth by the
Spitzer space telescope (Deming et al. 2007, Demory et al. 2007) indicates that the planet has an $8\,\mu m$ brightness 
temperature of $T=712\pm36\,{\rm K}$. If we adopt $T_{\rm eff}=3350\,{\rm K}$ for Gliese 436A, take
a zero albedo for the planet, and assume a uniform re-radiation of the orbit-averaged received flux from the entire planetary surface, we obtain a planetary effective temperature $T_{eq}=642\,{\rm K}$. The somewhat higher temperature implied by the secondary eclipse depth could arise from 
inefficient transport of heat to the night side of the planet, from a non-blackbody planetary 
emission spectrum, from tidal luminosity, or from a combination of the three.  The possibility that tidal luminosity is being observed is prompted by the orbital phase, $\phi=0.587$, of the secondary eclipse, which confirms that the orbital
eccentricity is alarmingly high (with a best-fit value of $e=0.150\pm 0.012$; Deming et al. 2007). Indeed, the planet and its primary
bring to mind an oversize alien version of Jupiter and Io.

The short orbital period and large eccentricity of Gliese 436b suggest that tidal circularization might be highly effective. To second order in eccentricity, the tidal luminosity of a
spin-synchronous planet (Peale \& Cassen 1978, Mardling 2007) is given by
\begin{equation}
{dE\over{dt}}={21\over{2}}{k_{2}\over{Q}}{G M_{\star}^{2} n R_{b}^{5} e^{2}\over{a^{6}}} \, ,
\end{equation}
where $k_2$ is the planetary tidal Love number, $a$ is the semi-major axis, $n$ is the orbital mean motion, and $Q$ is the planet's effective tidal dissipation parameter. Note that Gliese 436b cannot be fully spin synchronized, and because it presumably has no permanent quadropole moment, it should not reside in spin-orbit resonance. The pseudo-synchronization theory of Hut (1981; see also Goldreich \& Peale 1966) suggests $P_{\rm spin}=2.32 {\rm d}$, leading to a 19-day synodic period for the planet b. The analysis of Levrard et al. (2007), furthermore,
indicates that this spin asynchronicity of the planet will cause the tidal luminosity to exceed that given
by the above formula by a small amount. 

To date, several authors have investigated the Gliese 436 system with an eye toward obtaining a convincing explanation for planet b's large eccentricity. To establish context, it is useful to review these contributions.

\subsubsubsection{Solution 1: A large $Q$}

Jackson et al (2008a) have performed tidally-dissipated single-planet integrations backwards in time, and have shown that $Q_{b}/k_{2b} \approx 10^{6.6}$ and $Q_{\star}/k_{2\star} \approx 10^{5.5}$ can explain the high eccentricity of Gliese 436b. However, are such Q-values realistic?
The resonant configuration of the Jovian satellites implies $10^5<Q_{\rm J}<2\times10^6$, with the preferred value closer to the lower limit (Goldreich \& Soter 1966, Peale \& Greenberg 1980). Meyer \& Wisdom (2008) argue that $ Q_{\rm S} < 10^{5}$ whereas the circularization boundary for extrasolar planets (Wu 2003) suggests typical $Q$ values of order $3\times10^5$, allowing for significant variation from planet to planet in addition to significant overall uncertainty. Uranus and Neptune, which more closely resemble Gliese 436b in mass, composition, and radius, are a further factor of approximately ten times more dissipative.
Tittemore \& Wisdom (1989) have demonstrated that $Q_{\rm U} \lesssim 3.9 \times 10^{4}$, whereas in the case of Neptune, Banfield \& Murray (1992) find $1.2 \times 10^{4} < Q_{\rm N} < 3.3 \times 10^{5}$. Recent calculations of Zhang \& Hamilton (2008) suggest an even lower estimate: $9 \times 10^{3} < Q_{\rm N} < 3.6 \times 10^{4}$.

A theory for the origin of planetary $Q$ values (not to speak of a definitive understanding of dynamical tides) remains elusive, and so there is no a-priori reason to reject the hypothesis that Gliese 436b has a remarkably small internal dissipation. It should be noted, however, that for the particular case of Gliese 436b, $Q$ would need to be even higher than Jackson et al. (2008a)'s estimate of $Q_{b}/k_{2b} \approx 10^{6.6}$, given the likely breakdown of second-order equilibrium tidal theory at the very high initial eccentricities that they considered (Wisdom, 2008).

\subsubsubsection{Solution 2: A resonant perturber}

A natural solution to the high eccentricity problem is to invoke perturbations
from an as-yet undetected second planet in the system. Ribas et al. (2008) have suggested
that the perturbing body lies in a 2:1 mean motion resonance with the primary planet.
By constructing a 2-planet fit to the radial velocity data of Maness et al. (2007), they produced a
specific orbital model for the second planet, and predicted that changes in the eclipse depth and duration (arising from precession of Gliese 436b driven by the interaction with c should be
observable. 

Resonant perturbations contribute to the disturbing function with terms that contain mean longitudes, $\lambda$, and so they are generally capable of producing far larger transit timing 
variations than secular perturbations. Indeed, most discussions of the transit timing technique have
focused on perturbing planets in low-order mean motion resonances (e.g. Agol et al. 2005, Holman \& Murray 2005). In the  context of Gliese 436, the Ribas et al. (2008) model generates transit timing
variations measured in minutes, making their hypothesis eminently testable.

The $2.64\,{\rm d}$ period of planet b allows transit measurements to accumulate rapidly, and during the time that has elapsed since the Gillon et al. (2007) discovery, a number of high-precision observations of transit midpoint times have become publicly available. Alonso et al. (2008) analyzed the archival Spitzer photometry obtained via ``target of opportunity" observations in June 2007, and  derived $T_c=2454280.78191\pm0.00028\, {\rm HJD}$ (24 second accuracy). The same authors also analyzed their own ground-based observations of the March 8, 2008 transit, and derived $T_c=2454534.59584 \pm 0.00015\, {\rm HJD}$ (13 second accuracy).  By comparing these
transit times with Gillon et al (2007)'s discovery ephermeris of $T_c=2454222.616\pm0.001 {\rm HJD}$ (86 second accuracy), Alonso et al. (2008) then determined that there is only a $\sim0.5\%$ probability that a 5 $M_{\oplus}$ perturber in an exterior 2:1 mean motion resonance would have produced transit timing variations (TTV) smaller than those implied by the three midpoint times in their analysis. Recent work by Pont et al. (2008) has pushed the threshold for TTVs below 7 seconds, ruling out perturbers in 2:1 resonance down to lunar mass. Mardling (2008), furthermore, shows that semi-major axis decay produced by tidal dissipation will cause any 2:1 resonance between GL436b and a hypothetical companion to be short-lived. In particular, a perturber with the properties proposed by Ribas et al. (2008) passes through a 2:1 resonance on a timescale of $ \tau_{p} = 0.3(Q_{b}/10^{5})$ Gyr.

\subsubsubsection{Solution 3: A secular perturber}

In a recent paper, Tong and Zhou (2008) explored Gliese 436b's eccentricity modulation by both secular and resonant companions. Their approach adopted an octopole-order secular theory (see below) to identify hypothetical perturbers that can excite planet b's eccentricity (in the course of a single secular cycle) from $e_b=0$ to $e_b=0.16$, while simultaneously requiring that the system as a whole remain dynamically stable and that the perturber induce a radial velocity half amplitude less than the current threshold of detection (which they took to be $K<3\,{\rm m\, s^{-1}}$). Interestingly, they located a range of candidate perturbers that satisfied the constraints. Tong \& Zhou (2008) found, however, that when the effects of tidal dissipation were incorporated and the hypothetical two-planet systems were integrated forward in time, all eccentricity modulation was damped out and the orbits quickly circularized. As a part of their survey, Tong \& Zhou (2008) also specifically exclude the class of perturbing companion planets suggested by Maness et al. (2007), noting, for example that a putative perturber with $M_{c}=0.12 \, M_{\rm Jup}$, $P=14\,{\rm yr}$, and $e_c=0.6$ is capable of exciting an eccentricity, $e_{b_{max}}$, for Gl 436b to only $3\times10^{-6}$.

The aggresively effective orbital circularization observed by Tong \& Zhou (2008) led those authors to conclude that a companion planet (exerting either resonant or secular perturbations) that is consistent with the current corpus of Doppler velocity observations ``can not excite and maintain the significant eccentricity'' of Gl 436b. They argue that in order to adequately explain Gliese 436b's non-circular orbit, its tidal quality factor must be of order $Q_{b}/k_{2b}\sim6\times10^6$, a factor likely an order of magnitude larger than the values generally ascribed to gas giants (see however, Jackson, Barnes \& Greenberg (2008b) who argue that the upper limits on Jupiter's $Q$ may be significantly higher than generally believed, and see Matsumura, Takeda \& Rasio (2008) who present evidence of high Q for the aggregate of known extrasolar planets.)

In any case, the non-circular orbit of Gliese 436b constitutes a remarkable puzzle. The solution likely has interesting implications for our understanding of planetary structure, formation, and dynamics. The problem is rendered even more worthwhile by the fact that improved spectroscopic and photometric observations can be used to test the predictions inherent in a particular solution, as has been demonstrated already with the Ribas et al. (2008) hypothesis. This paper reports our attempt at a {\it dynamical} solution using currently published observational constraints. We investigate an alternate hypothesis that while the perturbations
are secular in nature, the eccentricity is not modulated, and the orbits of both planets are quasi-stationary. 
We present a semi-analytical survey which demonstrates that a range of well-defined apsidally aligned perturbers are individually capable of prolonging the effective circularization timescale of Gliese 436 b by more than a factor of five. We use numerical integrations to correct for planet-planet dynamics not captured by our secular theory, and present our final locus of predictions for the perturber properties in the form of a table. We confirm the long-term evolution for a select test case with a fully numerical scheme, and we discuss the possibilities for near-term observational confirmation. Finally, we discuss the possibility of a multiple-perturber scenario in the context of a modified Laplace-Lagrange secular theory and perform one proof-of-concept numerical integration.

\section{The Dynamical State of the System}

We first investigate the current dynamical state of a two-planet system. Throughout out analysis, we assume that the inclinations of perturbing planets with respect to Gliese 436b's orbit are dynamically negligible. This assumption may turn out to be overly restrictive, but we note that if the orbital angular momentum vector of Gliese 436b is found to be in near-alignment with the rotation axis of the parent star, then the possibility that the system is nearly co-planar is strengthened.  To date, however, there have been no reported observations of the Rossiter-McLaughlin effect (see e.g. Winn et al. 2005) during transits of Gliese 436b.

In the absence of mean motion resonances, the dynamics of a stable multi-planet system are dominated by secular interactions, with tidal forces also playing a significant role. In this scenario, any initial secular variations of the planets' orbital elements are damped out over $\sim 3$ tidal circularization timescales for the inner planet, given by
\begin{equation}
\tau_{b} = \left( \frac{G M_{\star} m_{b}}{a_{b}} \right) \left( \frac{e_{b}^{2}}{\dot{E}} \right)\, ,
\end{equation}
and the system approaches a quasi-stationary state. Once the fixed point is reached, the system begins to evolve exponentially towards a doubly circular state over a timescale that is considerably longer than than the circularization timescale (Mardling 2007). 

Correcting for the leading-order effects of general relativity (GR), which are crucial to this system, the orbit-averaged rates of change of eccentricities and longitudes of perihelia of the coplanar system are given by (Mardling \& Lin 2002; Mardling 2007)
\begin{equation}
\frac{de_{b}}{dt}=-\frac{15}{16} n_{b} e_{c}\left( \frac{m_{c}}{M_{\star}} \right) \left( \frac{a_{b}}{a_{c}} \right)^{4} \frac{\sin(\varpi_{b} - \varpi_{c})}{(1- e_{c}^{2})^{5/2}}\, ,
\end{equation}
\begin{equation}
\frac{de_{c}}{dt}=\frac{15}{16} n_{c} e_{b}\left( \frac{m_{b}}{M_{\star}} \right) \left( \frac{a_{b}}{a_{c}} \right)^{3} \frac{\sin(\varpi_{b} - \varpi_{c})}{(1- e_{c}^{2})^{2}}\, ,
\end{equation}
\begin{equation}
\frac{d \varpi_{b}}{dt}=\frac{3}{4} n_{b} \left( \frac{m_{c}}{M_{\star}} \right) \left( \frac{a_{b}}{a_{c}} \right)^{3}\frac{1}{(1- e_{c}^{2})^{3/2}} \left[ 1- \frac{5}{4} \left( \frac{a_{b}}{a_{c}} \right) \left( \frac{e_{c}}{e_{b}} \right) \frac{\cos(\varpi_{b} - \varpi_{c})}{1- e_{c}^{2}} \right]+\frac{3 n_{b}^{3}}{1-e_{b}^{2}} \left( \frac{a_{b}}{c} \right)^{2}\, ,
\end{equation}
and
\begin{equation}
\frac{d \varpi_{c}}{dt}=\frac{3}{4} n_{c} \left( \frac{m_{b}}{M_{\star}} \right) \left( \frac{a_{b}}{a_{c}} \right)^{2}\frac{1}{(1- e_{c}^{2})^{2}} \left[ 1- \frac{5}{4} \left( \frac{a_{b}}{a_{c}} \right) \left( \frac{e_{b}}{e_{c}} \right) \frac{(1+4e_{c}^{2})}{(1- e_{c}^{2})} \cos(\varpi_{b} - \varpi_{c}) \right]\, ,
\end{equation}
where $c$ is the speed of light, and $\varpi$ is the longitude of perihelion. The above set of equations is derived from the evolution of specific angular momentum and Runge-Lenz vectors utilizing a Legendre-polynomial expansion in terms of semi-major axis ratios $(a_{b}/a_{c})$ to octopole order (see Mardling \& Lin 2002 for the derivation), rather than the classical eccentricity expansion of the disturbing function (Laplace 1825,  Brouwer \& Clemence 1961). The model contains no constraints on the perturber's eccentricity, but it requires $(a_{b}/a_{c})$ to be small.
From equations (3 - 6), it is easy to see that there exist two apsidal fixed points: $(\varpi_{b} - \varpi_{c}) = 0$ and $(\varpi_{b} - \varpi_{c}) = \pi$. Each is stable for its own combination of orbital parameters, and libration around either can occur. Consequently, the system will tidally evolve towards an apsidally aligned fixed point if 
\begin{equation}
\frac{m_{c}}{m_{b}} > \frac{\sqrt{ a_{b}/a_{c} }}{(1-e_{c}^{2})(1+4(n_{b} a_{b}/c)^{2} (M_{\star}/m_{c})(a_{c}/a_{b})^{3}(1-e_{c}^{2})^{3}},
\end{equation}
and towards an anti-aligned state otherwise (Mardling, 2007). Note that our model treats the star as a point mass, thereby neglecting its tidal and rotational deformation's contribution to the apsidal precession rates of the planetary orbits.   While the rotational period of the star could not be determined from photometric observations (Butler et al 2004), if it is short, and the star's zonal gravitational harmonics are large, such terms could provide significant corrections to the equations presented here. Nonetheless, a fast rotation rate would be extremely uncharacteristic of the star's multi-Gyr age. An additional effect that we do not consider is precession arising from the quadropole moments of the planets b, and c. For ultra-short period planets with large radii, this effect can be important, but for Gliese 436b, it is small in comparison to the precession induced by the GR correction (see, e.g. Ragozzine \& Wolf 2008).

We can use our formulation to survey parameter space for regions where the perturbing planet might lie. While the age of the system is estimated to lie in the range of $\tau_{sys} \approx 1$Gyr $- 10$Gyr (Torres 2007), it almost certainly exceeds the few circularization timescales needed for the system to reach a fixed point. As a result, in the following analysis, we assume that the system is currently quasi-stationary, allowing us to restrict our search to fixed-point configurations. 

The following scheme was used to generate fixed point states of the system: the perturber's semi-major axis, eccentricity and mass were chosen randomly; equation (7) was invoked to determine whether the resulting fixed point is aligned or anti-aligned. By setting $d\varpi_{b}/dt = d\varpi_{c}/dt $, the eccentricity of Gliese 436b was determined. If the resulting eccentricity fell between $0.149$ and $0.151$, the configuration was kept. Subsequently, the generated system was tested against two constraints. First, to exclude grossly unstable systems, we required that the distance of closest approach of the two orbits was less than the radius of the larger Hill sphere, given by (Murray \& Dermott, 1999)
\begin{equation}
R_{H} = a(1-e) \left( \frac{m}{3M_{\star}} \right)^{1/3}.
\end{equation}
In practice, this criterion will admit some unstable systems; candidate systems that make relatively close approaches can be integrated numerically to verify stability.

 A planet with given values for period, mass, and eccentricity produces a radial velocity half-amplitude
\begin{equation}
K=({2 \pi G \over{P_c}})^{1/3} 
({M_{\rm c} \sin(i) \over{(M_{\star}+M_{\rm c})^{2/3}}}) 
{1\over{1-e_{c}^{2}}} \, .
\end{equation}
Conseqently, we required that the radial velocity half-amplitude produced by the perturber be less than 5m/s (a relatively generous criterion, even for perturbers of high eccentricity). This criterion rids the parameter space of readily detectable planets. Whenever one of our randomly generated systems satisfied our search criteria, it was numerically integrated forward in time, using the secular ODEs presented above, with the effects of tidal dissipation accounted for by including additional terms for the eccentricity and semi-major axis damping that arise from tidal dissipation in Gliese 436b
\begin{equation}
\frac{de_{b}}{dt}=-e_{b}\frac{21 \pi}{P_{b}} \frac{k_{b}}{Q_{b}} \frac{M_{\star}}{m_{b}} \left(\frac{R_{b}}{a_{b}}\right)^{5},  
\end{equation}
\begin{equation}
\frac{da_{b}}{dt}=-e_{b}^{2} a_{b} \frac{42 \pi}{P_{b}} \frac{k_{b}}{Q_{b}} \frac{M_{\star}}{m_{b}} \left(\frac{R_{b}}{a_{b}}\right)^{5}.  
\end{equation}
For each trial system, we can thus evaluate the system's new circularization timescale by noting the time, $t$, when Gliese 436b's initial eccentricity has decreased by a factor of $\mathit{e}$. To speed up the integration process, an artificially low $Q$-value of 300 was used. For the current purposes, we are interested less in the circularization timescale itself, but rather in the factor, $f$, by which the circularization of b is increased when the perturber is added. Consequently, we divide the numerically computed timescale by that of a secular integration with only Gliese 436b present.  

A total of 1625 model systems were generated and then integrated forward using the semi-analytic model. The results of this Monte-Carlo parameter survey are presented in Figures (1) through (4). Note that the perturbers at the anti-aligned fixed points (figures 3 and 4) generally have smaller periods than the perturbers in aligned configurations (figures 1 and 2). There is also a considerable difference in the factor by which the circularization time scale of planet b is increased. The anti-aligned perturbers provide an average increase factor of only $1.48 \pm 0.36$. The aligned perturbers on the other hand provide an average increase factor of $5.54 \pm 0.26$. The relatively high current eccentricity of GL 436b suggests a long circulization time, thus favoring the aligned perturbers. Within the aggregate of aligned perturbers, the damping timescale increase factor varies as well, peaking around $P_{c} \approx 37$d, $e_{c} \approx 0.58$, $M_{c} \approx 8_{\oplus}$ with an increase factor of $f=6.03$. A further attractive feature of the aligned configurations is that they form a very well-defined locus in ($P_c$, $e_c$, $M_c$) space. It is interesting to note that the region of parameter space occupied by viable perturbing companions is strongly influenced by the inclusion of general relativistic precession in Equation 5. With a purely Newtonian potential, apsidal alignment can only be maintained for perturbers with significantly longer orbital periods and significantly larger masses. This result is typical of many multiple-planet extrasolar systems, where one finds that the steady-state dynamical configuration has a delicate dependence on the inner planet precession rate (see, e.g. Adams \& Laughlin 2006a,b) 

\section{Numerical Confirmation}

The semi-analytical approach presented above is an effective method for conducting a parameter search for potential perturbing companions and the numerical integration of the secular equations with dissipation provides an efficient procedure for evaluating the resulting increase in orbital circularization time. The octopole order theory, however, is only approximate, and one would like to have a verification that does not incorporate the approximations. Indeed, to determine the exact fixed-point parameters of a hypothetical perturber, a fully numerical determination is much better suited. As a proof of concept, we explored twenty-seven representative cases, each containing an apsidally aligned perturber.

The determination of the fixed point parameters in the presence of the full equations of motion was performed by picking a perturber's period and mass, and subsequently performing many direct integrations while fine-tuning the perturber's eccentricity using a Newton-Raphson scheme until a combination of orbital elements that corresponds to a true fixed point was reached. The {\it Mercury6} integrator (Chambers, 1999) was used. The effects of general relativity were modeled by adding an extra term to the star's gravitational potential that takes the form
\begin{equation}
V_{GR}=-3 \left( \frac{G M_{\star}}{c \ r} \right)^{2},
\end{equation}
where $r$ is the distance from the planet to the star. The addition of the extra potential term yields the exact secular precession rate given by GR, while introducing a small error (of order $(m/M_{\star})^{2}$)in the planet's mean motion (Nobili \& Roxburgh, 1986). The so-called hybrid algorithm was used for these integrations. All of the integrations within the Newton-Raphson iteration  were short-term, never exceeding $10^{4}$ years. 

Using this numerical scheme, we determined the fixed point eccentricities of the potential perturbing companions, spaced with periods in equal increments between $P_{lower}=16\,{\rm d}$ and $P_{upper}=68\,{\rm d}$. In Table 1, we list the resulting systems (which are also plotted as a sequence of blue squares on Figures 1 and 2). The perturber's eccentricities can be approximated by a second order polynomial function: $e_{c} \approx - 0.277 + 2.94 \times 10^{-2} P_{c} - 2.05 \times 10^{-4} P_{c}^{2}$, where the period is measured in days (plotted as a dashed line on figure 2). It is clear that the inclusion of the full disturbing function produces some corrections to the octopole-order analytic solutions. Namely, secular theory tends to over-estimate the fixed-point eccentricities of the perturbers for this system.

Table 1 lists transit timing variations, $\Delta T_{\rm max}$ for the fixed point systems. The amplitude of the timing variations changes by a few seconds along this locus of models in ($P_c$, $e_c$, $M_c$) space. Much of this variation is due to the given configuration, residing close to, or in a high-order MMR. An extreme example is a $P_c=58\,{\rm d}$ perturber system, where the planets reside in a 22:1 resonance. Due to the perturber's very high fixed-point eccentricity, transit timing variations reach almost a minute. Hence, if our theory is correct, a given measurement of $\Delta t_{max}$ may provide an immediate hint to $P_c$, $e_c$, and $M_c$.

Our iteratively determined fixed-point models all appear to be indefinitely dynamically stable. By definition, at the fixed point, orbital parameter variations are reduced to the slow tidal circularization of both orbits, and so the planets never get into trouble. Interestingly, however, for eccentricities $e_c  \gtrsim 0.8$, we find that the octopole-order fixed-point solutions, when integrated using the full equations of motion, are largely unstable. When given access to the full disturbing function, these configurations usually experience eccentricity variations and librations in $\varpi_c-\varpi_b$ large enough to generate orbit crossings. The true fixed point configurations thus represent islands of stability within a larger sea, yielding a potentially important clue to the initial formation and evolution of the Gliese 436 planetary system. Indeed, if our model is confirmed by observation, then the question of system formation will represent an interesting line of inquiry.

Finally, we carried out a full numerical demonstration of the circularization timescale of the two-planet $P_c=40\,{\rm d}$ system listed on the thirteenth line of Table 1. This system lies near the region where the secular ODEs exhibit the largest factors of increase, $f$. To carry out the full integrations, we used a self-consistent numerical scheme that in additon to gravitational dynamics, takes into account the effects of tidal damping and spin-evolution (Eggleton et. al 1998, Mardling \& Lin 2002). In this simulation, the post-Newtonian gravitational force was computed in the same way as above. The tidal force takes the form
\begin{equation}
\mathbf{F}_{T} = -\left( \frac{6n (k_{2}/2)}{Q} \right) \left( \frac{M_{star}}{m} \right) \left( \frac{R}{a} \right)^{5} \left( \frac{a}{ | r | } \right )^{8} \times \left[ 3\left( \hat {\textbf{\textit{r}}} \cdot \dot{\textbf{\textit{r}}} \right) \hat{\textbf{\textit{r}}} + \left( \hat{\textbf{\textit{r}}} \times \dot{\textbf{\textit{r}}} - r  \mathbf{\Omega_{b}} \right) \times \hat{\textbf{\textit{r}}}   \right]
\end{equation}
and the spin evolution is governed by:
\begin{equation}
\dot{\mathbf{\Omega}}=\frac{m M_{\star}}{m + M_{\star}} \left( \frac{\textbf{\textit{r}} \times \mathbf{F}_{T}}{I} \right),
\end{equation}
where $\Omega$ is the spin vector and $I$ is the planet's moment of inertia. Note that, as before, tidal dissipation is applied only to Gliese 436b. As a result, Gliese 436b's $Q$-value absorbs all other dissipation in the system. Nonetheless, it is likely that the contribution from the perturber will be comparably small, since $dE/dt \propto (R/a)^{5}$, as will be the tides raised on the star where $dE/dt \propto (m/M_{\star})$. Here also, an artificially low $Q$-value of 10 (characteristic of Earth-like planets) was used in order to speed up the integration process. Integration was done using the Bulirsch-Stoer algorithm (Press et al. 1992) with a timestep accuracy parameter  $\eta=10^{-14}$. The moment of intertia was taken as $I/M_{\rm b}R_{\rm b}^2=0.254$ and the initial spin frequency was $\Omega=2.3\,{\rm d}$. As we have done throughout the paper, we use $P_{b}=2.643\,{\rm d}, e_{b}=0.15,$ and $m_{b}=0.0711 M_{J}$.

Two scenarios were computed: a tidally dissipated Gliese 436 system with and without a fixed-point $P_{c}=40\,{\rm d}$ perturber. Figure (5) shows the eccentricity evolutions of the perturbed and unperturbed Gliese 436b's as well as the eccentricity evolution of the perturber. The results of this integration are in good agreement with the semi-analytical integrations discussed in section 2. By measuring the time at which Gliese 436b's eccentricity dropped by a factor of $e$ in the perturbed and unperturbed solutions, we find that the $P_{c}=40\,{\rm d}$ perturber provides a $f \sim 5.3$ times increase in the  circularization timescale. Figure (6) shows the difference of perihelia,  $(\varpi_{b} - \varpi_{c})$ as a function of time. Linear secular theory predicts that once the fixed point is reached, the system does not deviate away from it at all. In the full numerical integration, we observe some low-amplitude libration, which is associated with passage of Gliese 436b through high-order MMR's (namely 16:1, 17:1, 18:1, and 19:1 as labeled on figures 5 and 6) .

\section{Many-Planet Fixed Points}
Up to this point, we have only considered a scenario where the Gliese 436b's eccentricity is maintained by a single perturber. While a two-planet scenario is the simplest one, it is certainly possible to obtain solutions where three or more planets are all in a quasi-stationary configuration. In this case, to fully take advantage of the energy reservoir of the system, all consecutive pairs of planets should be dynamically coupled. Let us now examine fixed point systems in the context of a modified classical Laplace-Lagrange secular theory. 

To second order in eccentricities and first order in masses, the averaged co-planar disturbing function of planet $i$ can be written as
\begin{equation}
\left< \mathcal{R}_{i}^{(sec)} \right>=n_{i}a_{i}^{2}\left[\frac{1}{2} A_{ii} e_{i}^{2} + 
\sum_{j=1,j\neq{i}}^{N}A_{ij}e_{i}e_{j}\cos(\varpi_{i}-\varpi_{j})\right].
\end{equation}
The constant coefficients $A_{ij}$ form a N $\times$ N matrix, which renders the dynamics of the system (Murray \& Dermott, 1999). Here, the effects of GR are accounted for by introducing a leading-order correction to the diagonal elements of the $\bold{A}$ matrix (for details on this standard modification, see Adams \& Laughlin, 2006). As a result, the elements take the form:
\begin{equation}
A_{ii}=\frac{n_{i}}{4}\sum_{j=1,j\neq{i}}^{N}\frac{ {m_{j}} } {M_{\star}+m_{i}}\alpha_{ij} \bar{\alpha}_{ij} b_{3/2}^{(1)}(\alpha_{ij})+3n_{i}\frac{GM_{\star}}{c^{2}a_{i}},
\end{equation}
\begin{equation}
A_{ij}=-\frac{n_{i}}{4}\sum_{j=1,j\neq{i}}^{N}\frac{ {m_{j}} } {M_{\star}+m_{i}}\alpha_{ij} \bar{\alpha}_{ij} b_{3/2}^{(2)}(\alpha_{ij}),
\end{equation}
where $\alpha_{ij} = a_{i}/a_{j}$ if $( a_{i} < a_{j})$; $ a_{j}/a_{i}$ if $(a_{j} < a_{i})$, $\bar{\alpha}_{ij} = \alpha_{ij}$ if $( a_{i} < a_{j})$; $ 1$ if $(a_{j} < a_{i})$, $b_{3/2}^{(1)}(\alpha_{ij})$ \& $b_{3/2}^{(2)}(\alpha_{ij})$ are Laplace coefficients of first and second kind respectively.
The lowest-order Lagrange's equations of motion can be written as
\begin{equation}
\frac{de_i}{dt} = -\frac{1}{n_i a_i^2 e_i} \frac{\partial \mathcal{R}_{i}}{\partial \varpi_i} = \sum_{j=1,j\neq{i}}^{3}A_{ij}e_{j}\sin(\varpi_{i}-\varpi_{j})
\end{equation}
\begin{equation}
\frac{d\varpi_i}{dt} = \frac{1}{n_i a_i^2 e_i} \frac{\partial \mathcal{R}_{i}}{\partial e_i} = A_{ii} + \sum_{j=1,j\neq{i}}^{3}A_{ij}\frac{e_{j}}{e_i}\cos(\varpi_{i}-\varpi_{j}).
\end{equation}
From these expression, it can be seen that $de/dt$ vanishes when the orbits are apsidally aligned and anti-aligned. Since all precession rates of a fixed point system are equal, given the masses and semi-major axes of the perturbing planets, we can solve for their fixed-point eccentricities. Care must be taken in this approach however, since Laplace-Lagrange theory only applies in the limit of small eccentricities\footnote{The issue of small eccentricities can be forestalled by expanding the secular disturbing function to higher order in $e$.}.
 
Let us now consider a single proof-of-concept calculation. Setting the masses of the two perturbers at 22$m_{\oplus}$ each, we allowed the semi-major axes to be generated randomly, and then solved for the perturbers' aligned fixed-point eccentricities. We then used the numerical scheme discussed above to integrate the full equations of motion forward in time. The orbital elements of this configuration are presented in table 2, whereas figure 7 shows the system's evolution towards a triply-circular state. Since the eccentricities of this configuration are somewhat higher than what is required for Laplace-Lagrange theory to reach high levels of precision, it is not surprising that the analytically obtained fixed point is only approximately stationary, and some eccentricity modulation in the numerical integration is observed. 

In principle, a similar survey to that described in section 2 can be performed. However, as more planets are added, the parameter space, available for exploration, becomes overwhelmingly large. Thus, such a search would be most effective in case a perturbing planet is already found, but is clearly not at a single-perturber fixed point. Indeed, such a detection may indirectly hint to the presence of additional, dynamically significant, bodies in the system.

\section{Discussion}

In this paper, we put forward the hypothesis that the Gliese 436 system has an additional, as-yet undetected companion, Gliese 436c, that is {\it not} in a low-order mean motion resonance with Gliese 436b, but is at a secular fixed point. An important role of such a perturber would be to increase the tidal circulization timescale of the system and thus help explain the unusually high eccentricity, $e=0.15$, of Gliese 436b. 

Using the dissipated secular theory developed by Mardling (2007), we surveyed the parameter space of allowed perturbers, restricting our domain to fixed-point configurations. This survey revealed that apsidally aligned, fixed-point systems can provide significantly longer circularization times than the anti-aligned states, and are thus preferred. Furthermore, the population of possible stationary, apsidally aligned, perturbers forms a clear and well-defined locus in ($P_c\,$,$M_c\,$,$e_c\,$) space. This sequence is accompanied by the amplitude of transit timing variations, $\Delta T_{\rm max}$, listed in Table 1. As a check of the validity of our semi-analytic theory, we verified that the full disturbing function produces only small corrections to the orbital elements of the fixed-point configurations derived from octopole theory. In addition, we performed one self-consistent numerical integration of the circularization a hypothetical $P_{c}=40\,{\rm d}$ perturber. The results are in good agreement with the octopole theory, allowing us to verify that a factor, $f=5.3$ increase in circularization timescale is robust. 

A satisfactory theory makes predictions that are not only testable, but {\it readily} testable. The presence of Glise 436c, with the orbital and physical properties that we're proposing would fall into the readily testable category.

We make several predictions. {\bf 1} If a secular perturber exists and holds significant responsibility for the upkeep of Gliese 436b's eccentricity, we believe that it will fall on the  loci shown in Figures 1 and 2 and delineated in Table 1. {\bf 2} Our theory gives a roughly sixfold increase in the circularization timescale derived by Tong \& Zhou (2008). If we adopt $Q=300,000$, consistent with the upper bound on Banfield \& Murray's (1992) estimate of Neptune's $Q$-value, and near the lower bound on Jupiter's $Q$-value, then the effective (fixed-point) circularization timescale becomes $\tau \sim 8\,{\rm Gyr}$, a value that modestly exceeds the likely age of the star. In addition, the system may also require up to $\sim$ 3 single-planet circularization timescales to arrive at the fixed point. As a result, for the first time, we don't need to invoke an uncomfortably high Q-value to explain the observed eccentricity, but if our theory is correct then Gliese 436b's effective internal dissipation is a factor of 10 times less than Uranus and is near the lower end of Neptune's range, suggesting a fundamental structural difference between the ``hot Neptune" and the slightly less massive solar system ``ice giants". {\bf 3} We predict that Gliese 436b will show {\it no} measurable tidal heating. On the surface, this appears inconsistent with Deming et al 2007's measurement of  $T=712\pm36\,{\rm K}$ for the planet's $8\,\mu m$ brightness temperature. If our theory is correct, the apparent temperature excess over the $T_{eq}=642\,{\rm K}$ planetary effective temperature will arise from a non-blackbody emission spectrum, weather variability, measurment uncertainty, or a combination of the three.

In the previous section, we reviewed the possibility of more than one perturber maintaining Gliese 436b's eccentricity, utilizing modified Laplace-Lagrange secular theory, and performed one numerical integration as an example. In light of this possibility, we note that observational detection of a companion \textit{not} in a 2-planet fixed configuration may indirectly point at the presence of additional planets in the system. 

What are the prospects for near term observational detection of Gliese 436 c? During the two years following the radial velocity analysis of Maness et al. (2007) the Gliese 436 system has been placed under intensive radial velocity surveillance. At the IAU 253 symposium in May 2008, the Geneva Planet Search team (Lovis et al. 2008) showed several dozen new high-precision Doppler measurements, which indicated that the possible long-term linear trend noted by Maness et al. (2007) has either abated, or was an artifact of the limited observational baseline. A residuals periodogram of the Geneva team's velocity data showed a number of peaks in the $10\,{\rm d}<P<100\,{\rm d}$ range, several of which appeared tall enough to correspond to a perturber with the properties required by our theory.

The EPOXI mission (Deming et al. 2008) held Gliese 436 under continuous photometric observation for 27 days during May 2008. The full results of their investigation have not yet been published, but the EPOXI transit timing precision for Gliese 436b is expected to be of the order of a few tens of seconds (Deming 2009). At this level of precision, if our hypothesis of a fixed point perturber is correct, we predict that {\it no} transit timing variations will be observed by EPOXI. Thus far, our hypothesis has been consistent with the lack of transit timing variations above $\Delta t \sim 45$ seconds reported by Coughlin et al (2008). If transit timing variations do exist, their maximum amplitude will lie in the neighborhood of $\Delta t \sim 1$ to $\Delta t \sim 10$ seconds, and their maximum amplitude will give a strong indication of the perturber's orbital properties, as interpolated from Table 1. 

Recent improvements in high-cadence, high-precision ground-based photometry have improved the measurement of transit midpoints to within $\Delta t \sim 5$ seconds (see e.g. Johnson et al. 2008). If this level of precision can be repeatedly applied to Gliese 436, then there is a significant chance that the presence of a perturbing planet can be inferred. In addition, the predicted radial velocity half-amplitudes, $K$, of the perturber are generally well within the current threshold of detection, which should enable a ready confirmation or falsification of our hypothesis.

We thank Fred Adams, Dan Fabrycky, Rosemary Mardling, Stan Peale, David Stevenson, Darin Ragozzine, and Drake Deming for useful discussions. 

This research is based in part upon work supported by
the National Science Foundation CAREER program under Grant No. 0449986
(GL), NSF AST-0307493 Grant (SV) and by NASA through the Ames Astrobiology Institute under cooperative agreement NNA06CB31A.

\newpage 

\clearpage

\begin{table}
\begin{center}
\caption{Potential Fixed Point Perturbing Companions to Gliese 436b}
\begin{tabular}{ccccccccccccccc}
\tableline\tableline
TTV (s) & K (m/s) & P (days) & m ($M_{J}$) & e  \\
\tableline
2.46 & 4.39 & 16 & 3.06298 $\times 10^{-2}$ & 0.166647 \\
1.66 & 4.92 & 18 & 3.5136 $\times 10^{-2}$ & 0.208449 \\
1.12 & 3.87 & 20 & 2.85267 $\times 10^{-2}$ & 0.214894 \\
0.8 & 3.09 & 22 & 2.33118 $\times 10^{-2}$ & 0.229319 \\
0.11 & 3.22 & 24 & 2.43739 $\times 10^{-2}$ & 0.280283 \\
1.46 & 4.05 & 26 & 3.01091 $\times 10^{-2}$ & 0.344026 \\
1.73 & 4.88 & 28 & 3.56371 $\times 10^{-2}$ & 0.393356 \\
1.13 & 3.43 & 30 & 2.51117 $\times 10^{-2}$ & 0.415769 \\
0.88 & 2.72 & 32 & 1.96387 $\times 10^{-2}$ & 0.447974 \\
2.27 & 4.24 & 34 & 2.974 $\times 10^{-2}$ & 0.488534 \\
0.95 & 3.79 & 36 & 2.61039 $\times 10^{-2}$ & 0.517317 \\
1.37 & 3.43 & 38 & 2.30358 $\times 10^{-2}$ & 0.545275 \\
2.65 & 4.07 & 40 & 2.68 $\times 10^{-2}$ & 0.56866 \\
4.19 & 4.73 & 42 & 3.05687 $\times 10^{-2}$ & 0.589384 \\
2.35 & 3.68 & 44 & 2.30875 $\times 10^{-2}$ & 0.613239 \\
2.59 & 4.44 & 46 & 2.74619 $\times 10^{-2}$ & 0.627729 \\
2.77 & 3.51 & 48 & 2.08942 $\times 10^{-2}$ & 0.652055 \\
9.72 & 3.65 & 50 & 2.07423 $\times 10^{-2}$ & 0.676771 \\
2.77 & 4.44 & 52 & 2.56081 $\times 10^{-2}$ & 0.675824 \\
2.48 & 4.01 & 54 & 2.24914 $\times 10^{-2}$ & 0.691682 \\
4.35 & 3.68 & 56 & 1.99315 $\times 10^{-2}$ & 0.709095 \\
49.59 & 5.23 & 58 & 2.2408 $\times 10^{-2}$ & 0.781977 \\
2.88 & 3.08 & 60 & 1.57439 $\times 10^{-2}$ & 0.735554 \\
6.41 & 3.76 & 62 & 2.00128 $\times 10^{-2}$ & 0.726411 \\
3.54 & 4.34 & 64 & 2.18008 $\times 10^{-2}$ & 0.747703 \\
5.93 & 3.40 & 66 & 1.82263 $\times 10^{-2}$ & 0.731106 \\
4.15 & 4.19 & 68 & 2.01049 $\times 10^{-2}$ & 0.766115 \\
\tableline
\end{tabular}
\end{center}
\end{table}

\begin{table}
\begin{center}
\caption{An example 3-planet  fixed point configuration}
\begin{tabular}{ccccccccccccccc}
\tableline\tableline
Planet & P (days) & m ($M_J$) & e & $\varpi$ (deg) \\
\tableline
b & 2.64 & 7.11$\times 10^{-2}$ & 0.15 & 0.0\\
c & 14.24 & 6.92 $\times 10^{-2}$ & 0.139214 & 0.0\\
d & 27.33 & 6.92 $\times 10^{-2}$ & 0.152183 & 0.0\\
\tableline
\end{tabular}
\end{center}
\end{table}

\clearpage

\begin{figure*}[t]
\centering
\includegraphics[width=1.0\textwidth]{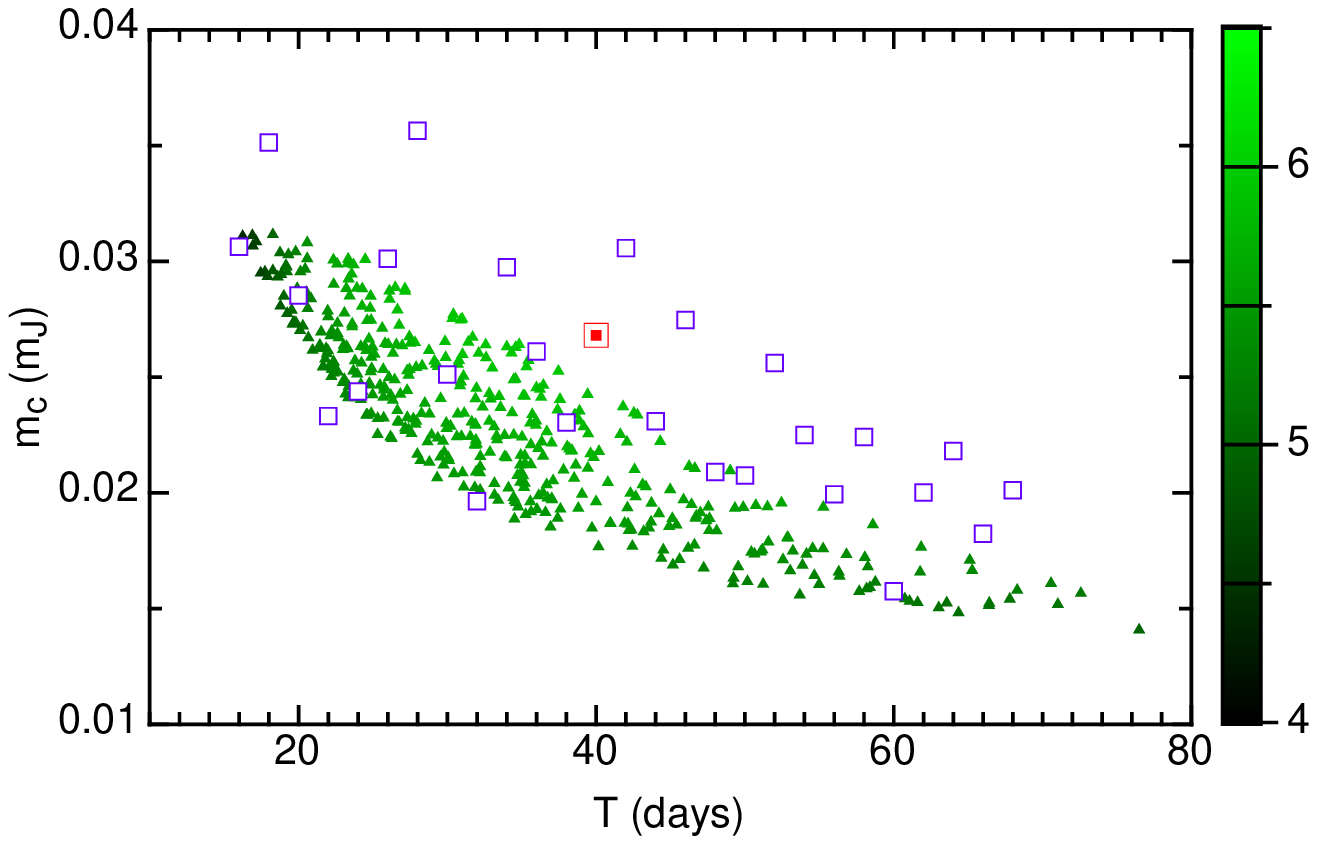}
\caption{Distribution of masses and orbital periods for potential Gliese 436 c candidates occupying apsidally aligned fixed points relative to the known transiting planet Gliese 436b ({\it small triangles}). Color indicates the factor of increase in the orbital circularization time for Gliese 436b in comparison to the case where no perturber is present. The color scale runs from black (4 times increase in time scale) to light green (factor of 6.5 increase in circularization time). The aligned perturbers produce significant increases in orbital circularization time, and hence are thus favored as an explanation for the substantial orbital eccentricity of Gliese 436b. Aligned fixed-point models obtained using the full equations of motion are shown as open blue squares. The red dot signifies the configuration that was used for initial conditions of the numerical integration of orbital decay.}
\end{figure*}

\begin{figure*}[t]
\centering
\includegraphics[width=1.0\textwidth]{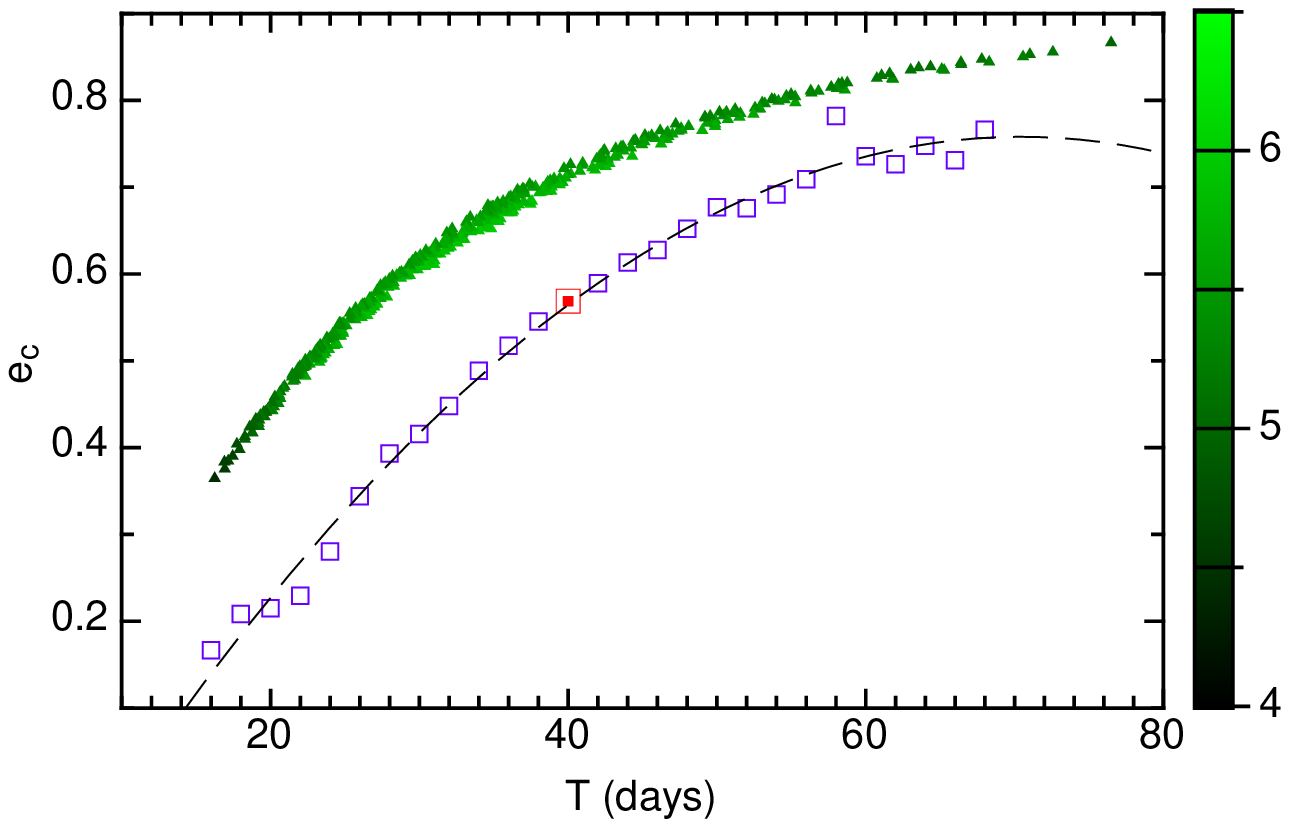}
\caption{Distribution of eccentricities and orbital periods for potential Gliese 436 c candidates occupying apsidally aligned fixed points relative to the known transiting planet Gliese 436b ({\it small triangles}). Color indicates the factor of increase in the orbital circularization time for Gliese 436b in comparison to the case where no perturber is present. The color scale runs from black (4 times increase in time scale) to light green (factor of 6.5 increase in circularization time). Aligned fixed-point models determined using the full equations of motion are shown as open blue squares. The red dot signifies the configuration that was used for initial conditions of the numerical integration of orbital decay. The dashed line corresponds to a best-fit second-order polynomial function (see section 3).}
\end{figure*}

\begin{figure*}[t]
\centering
\includegraphics[width=1.0\textwidth]{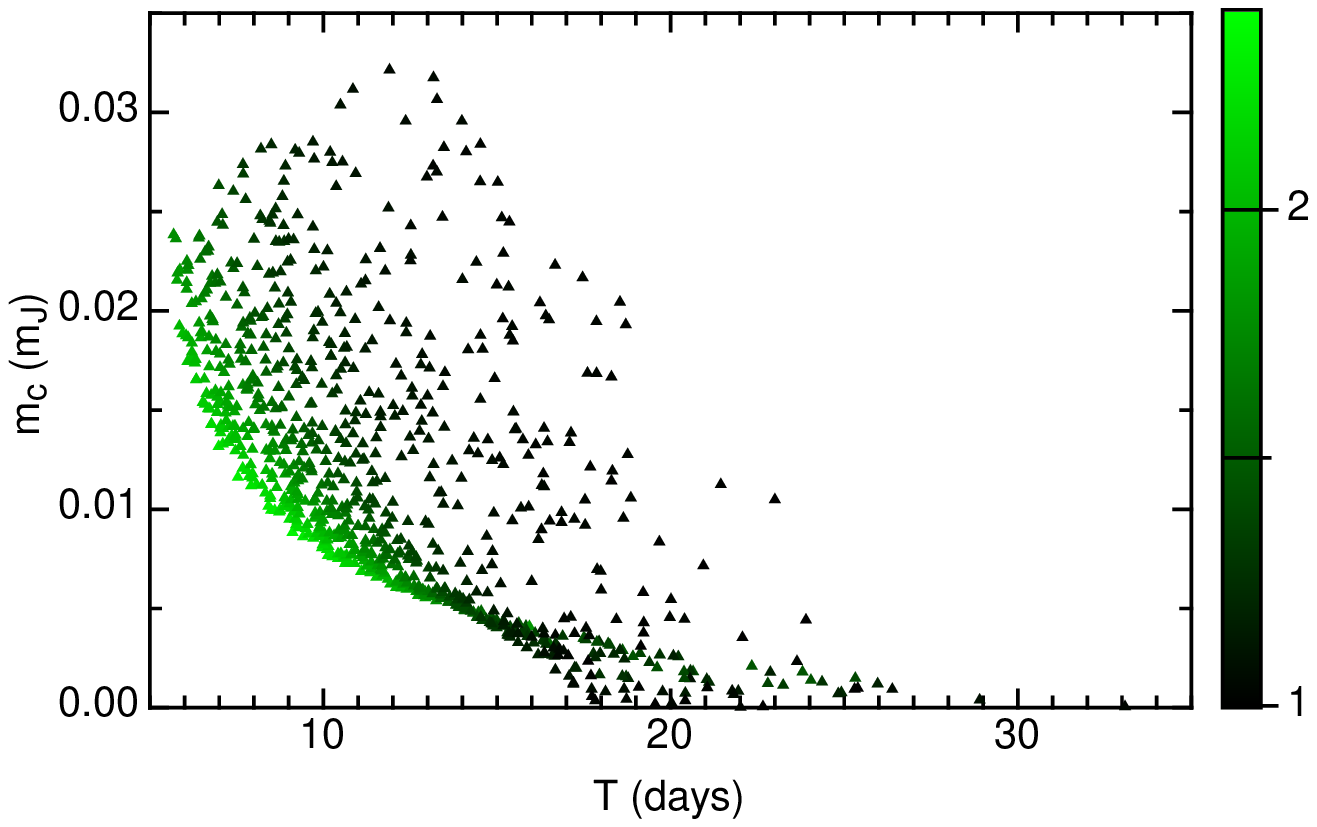}
\caption{Distribution of masses and orbital periods for potential Gliese 436 c candidates occupying apsidally anti-aligned fixed points relative to the known transiting planet Gliese 436b ({\it small triangles}). Color indicates the factor of increase in the orbital circularization time for Gliese 436b in comparison to the case where no perturber is present. The color scale runs from black (no increase in time scale) to light green (factor of 3 increase in circularization time). The anti-aligned perturbers produce only modest increases in orbital circularization time, and hence are thus less favored to explain the substantial orbital eccentricity of Gliese 436b.}
\end{figure*}

\begin{figure*}[t]
\centering
\includegraphics[width=1.0\textwidth]{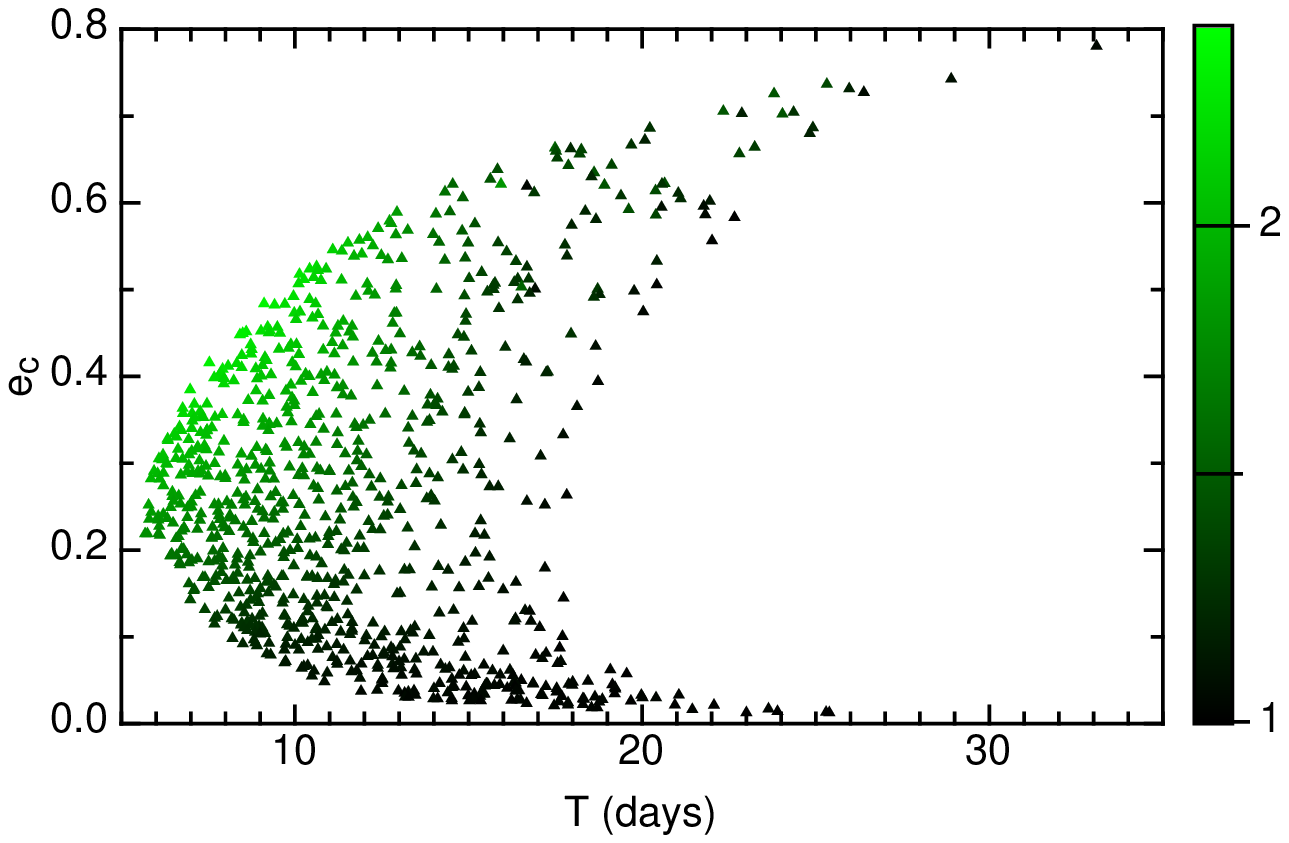}
\caption{Distribution of eccentricities and orbital periods for potential Gliese 436 c candidates occupying apsidally anti-aligned fixed points relative to the known transiting planet Gliese 436b ({\it small triangles}). Color indicates the factor of increase in the orbital circularization time for Gliese 436b in comparison to the case where no perturber is present. The color scale runs from black (no increase in time scale) to light green (factor of 3 increase in circularization time).}
\end{figure*}

\begin{figure*}[t]
\centering
\includegraphics[width=1.0\textwidth]{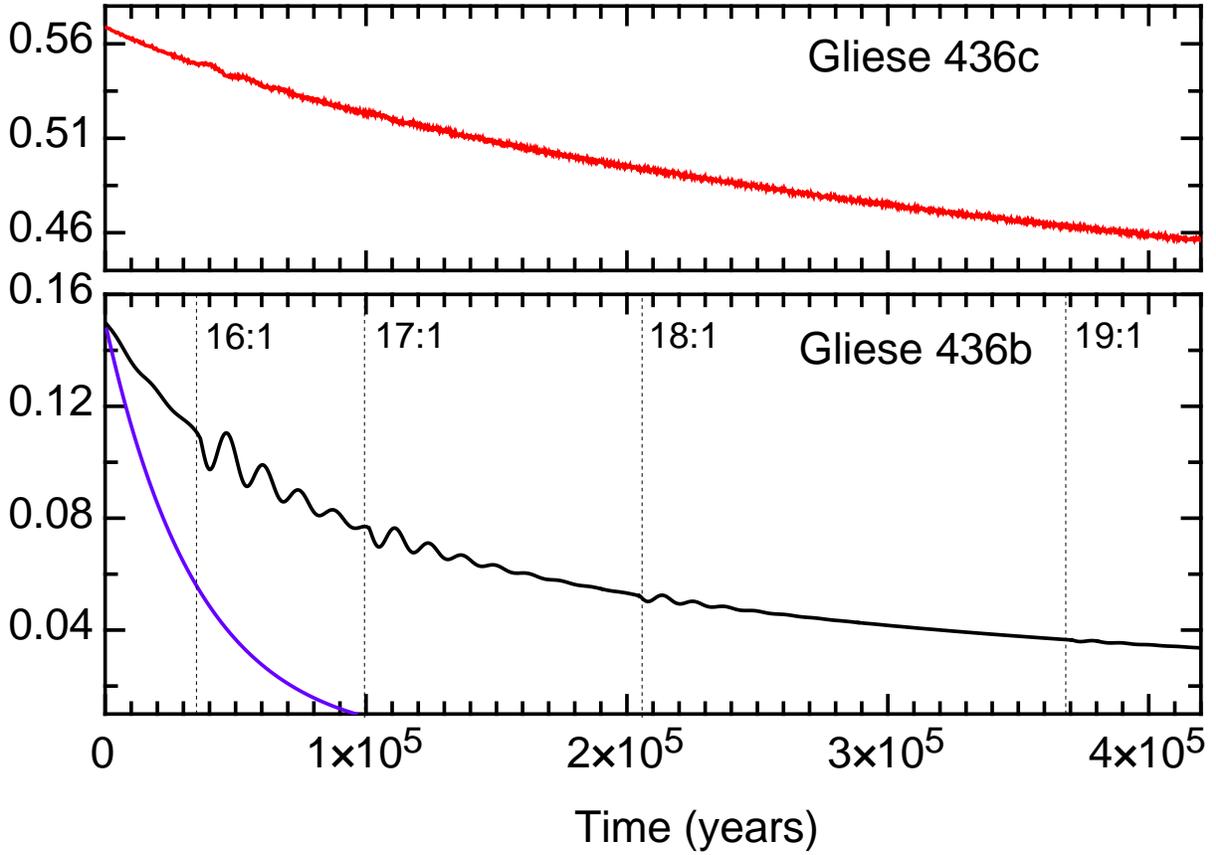}
\caption{Numerically computed eccentricity evolution of the tidally dissipated Gliese 436 system, starting from a numerically determined aligned fixed point. The top plot corresponds to the eccentricity of a $P=40\,{\rm d}$ perturber. The black curve in the bottom plot corresponds to the eccentricity evolution of Gliese 436b, in presence of the perturber. The blue curve in the bottom plot corresponds to the scenario where Gliese 436b is alone.  For the integrations, $Q=10$ was used. }
\end{figure*}

\begin{figure*}[t]
\centering
\includegraphics[width=1.0\textwidth]{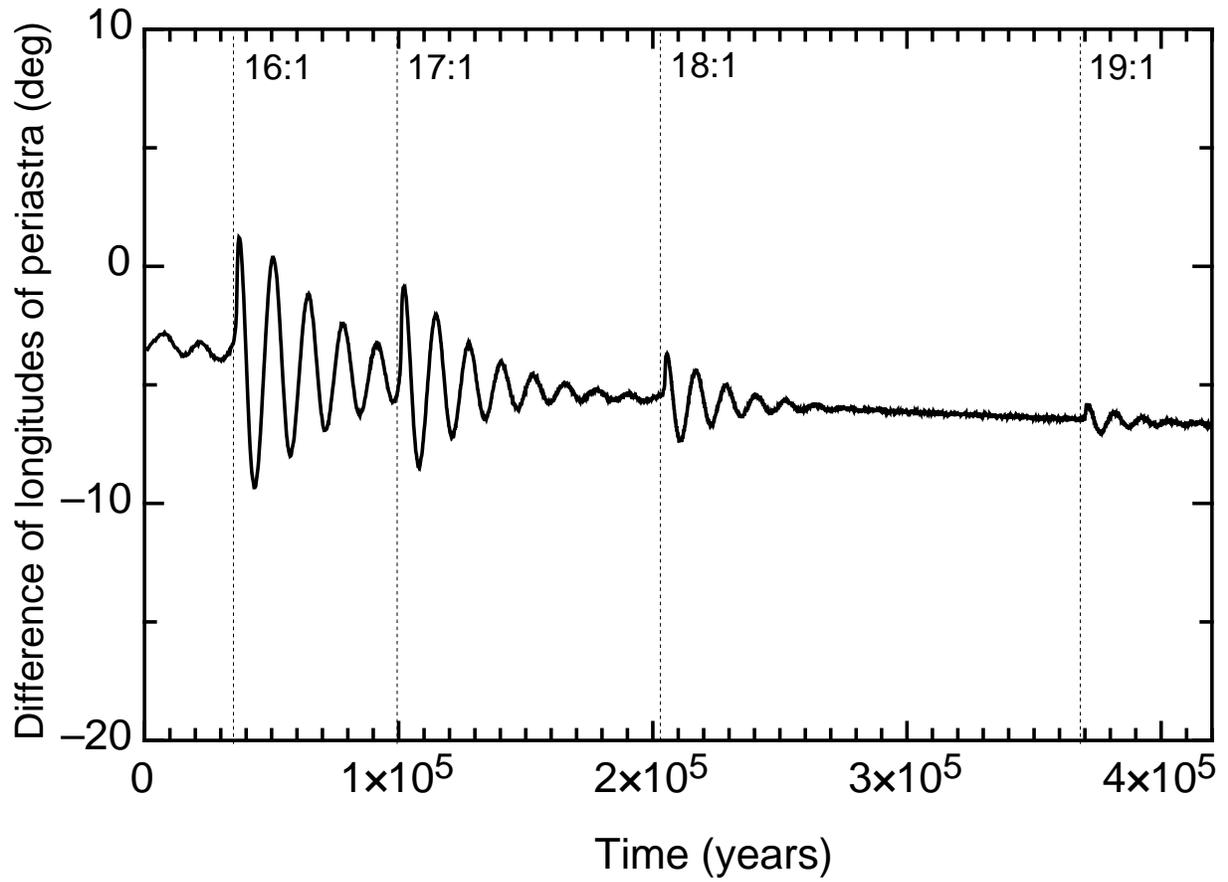}
\caption{The difference between apsidal lines of Gliese 436b and c for the numerically computed tidally dissipated eccentricity evolution shown in Figure 5.}
\end{figure*}

\begin{figure*}[t]
\centering
\includegraphics[width=1.0\textwidth]{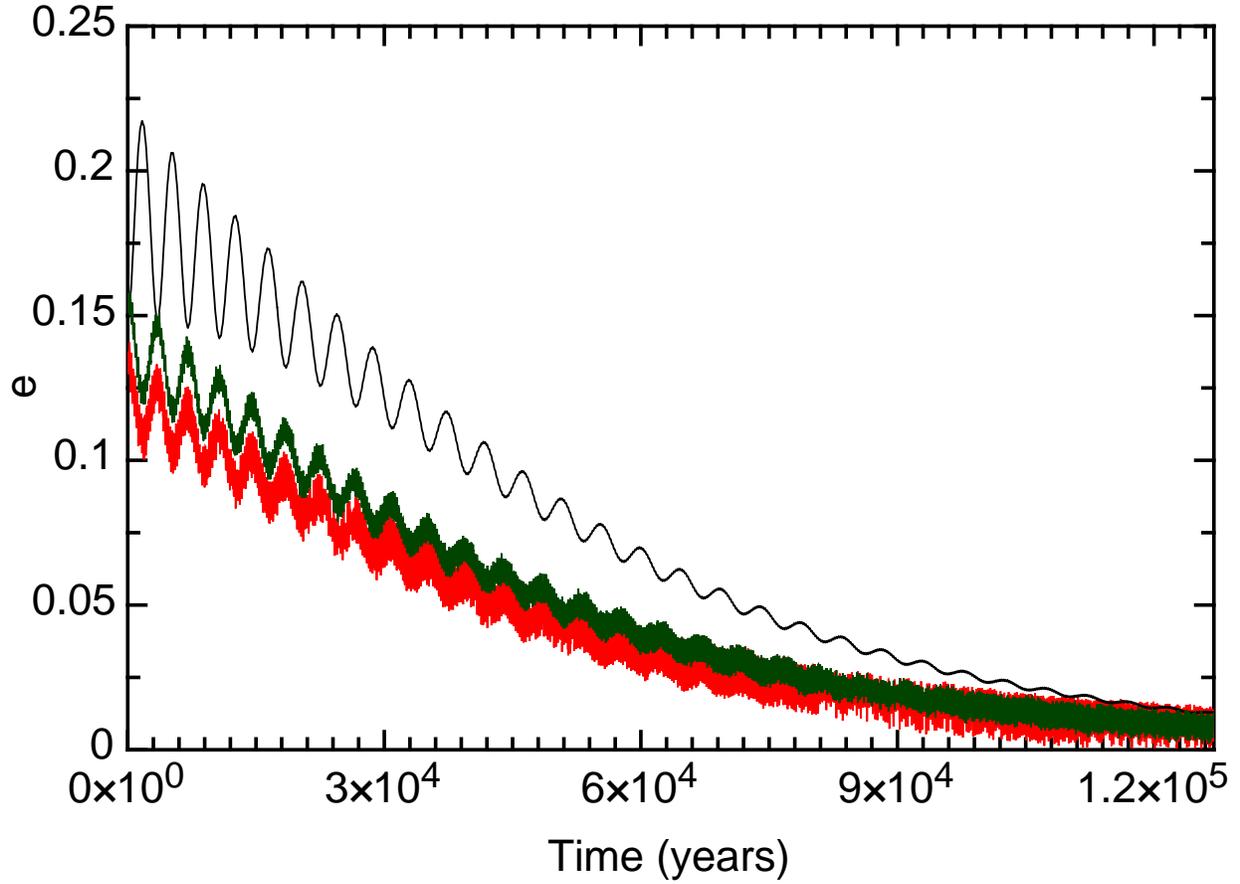}
\caption{Numerically computed eccentricity evolution of the tidally dissipated Gliese 436 system with 3 planets. The starting conditions were computed analytically, and are listed in Table 2. The black, red, and green curves correspond to planets b, c, and d respectively. For the integrations, $Q=8$ was used. }
\end{figure*}

\end{document}